\documentclass[twocolumn,prl,tightenlines,superscriptaddress,showpacs]{revtex4-1}

\usepackage{amsmath}
\usepackage{amssymb,amsfonts,latexsym}
\usepackage{bm}
\usepackage[mathcal]{euscript}
\usepackage{graphicx}
\usepackage{epsfig}
\usepackage{color}
\usepackage{xfrac}
\usepackage{mathrsfs}

\usepackage{hyperref}
\usepackage{url}

\begin{document}

\title{Condensation and Synchronization in Aligning Chiral Active Matter}

\author{Yujia Wang}
\affiliation{Center for Soft Condensed Matter Physics and Interdisciplinary Research, Soochow University, Suzhou 215006, China}

\author{Bruno Vent\'{e}jou}
\affiliation{Service de Physique de l'Etat Condens\'e, CEA, CNRS Universit\'e Paris-Saclay, CEA-Saclay, 91191 Gif-sur-Yvette, France}

\author{Hugues Chat\'{e}}
\affiliation{Service de Physique de l'Etat Condens\'e, CEA, CNRS Universit\'e Paris-Saclay, CEA-Saclay, 91191 Gif-sur-Yvette, France}
\affiliation{Computational Science Research Center, Beijing 100193, China}
\affiliation{Sorbonne Universit\'e, CNRS, Laboratoire de Physique Th\'eorique de la Mati\`ere Condens\'ee, 75005 Paris, France}

\author{Xia-qing Shi}
\affiliation{Center for Soft Condensed Matter Physics and Interdisciplinary Research, Soochow University, Suzhou 215006, China}

\date{\today}

\begin{abstract}
We show that spontaneous density segregation in dense systems of aligning 
circle swimmers is a condensation phenomenon at odds with the phase separation scenarios 
usually observed in two-dimensional active matter.
The condensates, which take the form of vortices or rotating polar packets,  
can absorb a finite fraction of the particles in the system, and keep a finite or slowly growing size as their mass increases. 
Our results are obtained both at particle and continuous levels.
We consider both ferromagnetic and nematic alignment, and both identical
and disordered chiralities. 
Condensation implies synchronization, even though our systems are in 2D and bear strictly local interactions.
We propose a phenomenological theory based on observed mechanisms that accounts qualitatively for our results. 
\end{abstract}

\maketitle

Spontaneous segregation of active matter into dense and sparse domains is ubiquitous.
In systems with local interactions, it is usually well described as phase separation,
and occurs not only in scalar active matter (``motility-induced phase separation'' \cite{cates2015motility}), 
but also in vectorial, aligning systems. 
This is in particular the generic situation for the simple but important case of
self-propelled particles locally aligning their velocities against some noise \cite{chate2020dry}. 
In such dry aligning active matter, the order-disorder transition is not direct,
and the homogeneous orientationally-ordered liquid
is generically separated from disorder by a coexistence phase in which dense ordered regions evolve
in a remaining vapor.

Like in equilibrium, a key feature of phase separation in active matter is that the dense phase occupies 
a finite fraction of the system's volume, either as a single domain or in the form of quantized
micro-phases, i.e. dense objects with a finite typical size~\footnote{In systems with long-range interactions, however, such as those governed by chemotaxis, condensation can occur (chemotactic collapse). See, e.g., \cite{saha2014clusters,pohl2014dynamic}}.

Chiral active matter~\footnote{see \cite{Liebchen_2022} for a short recent review and references therein.}, 
currently under intense scrutiny, seems to be no exception. Whether made of spinning particles or circle
swimmers, density segregation has been reported in numerous works 
in two dimensions (2D), with most frequently micro-phase-like separation 
\cite{schwarz2012phase,denk2016active,liebchen2016pattern,huang2020dynamical,ventejou2021susceptibility,moore2021chiral,scholz2021surfactants,massana-cid2021arrested,hokmabad2022spontaneously,ma2022dynamical}, but also
macro-phase separation \cite{levis2019activity,kruk2020traveling,tan2022odd,caporusso2024phase}, and mixed or hard to classify cases \cite{liebchen2017collective,levis2019simultaneous,Kreienkamp_2022,ai2023spontaneous,semwal2024macro}



\begin{figure}[b!]
\includegraphics[width=\columnwidth]{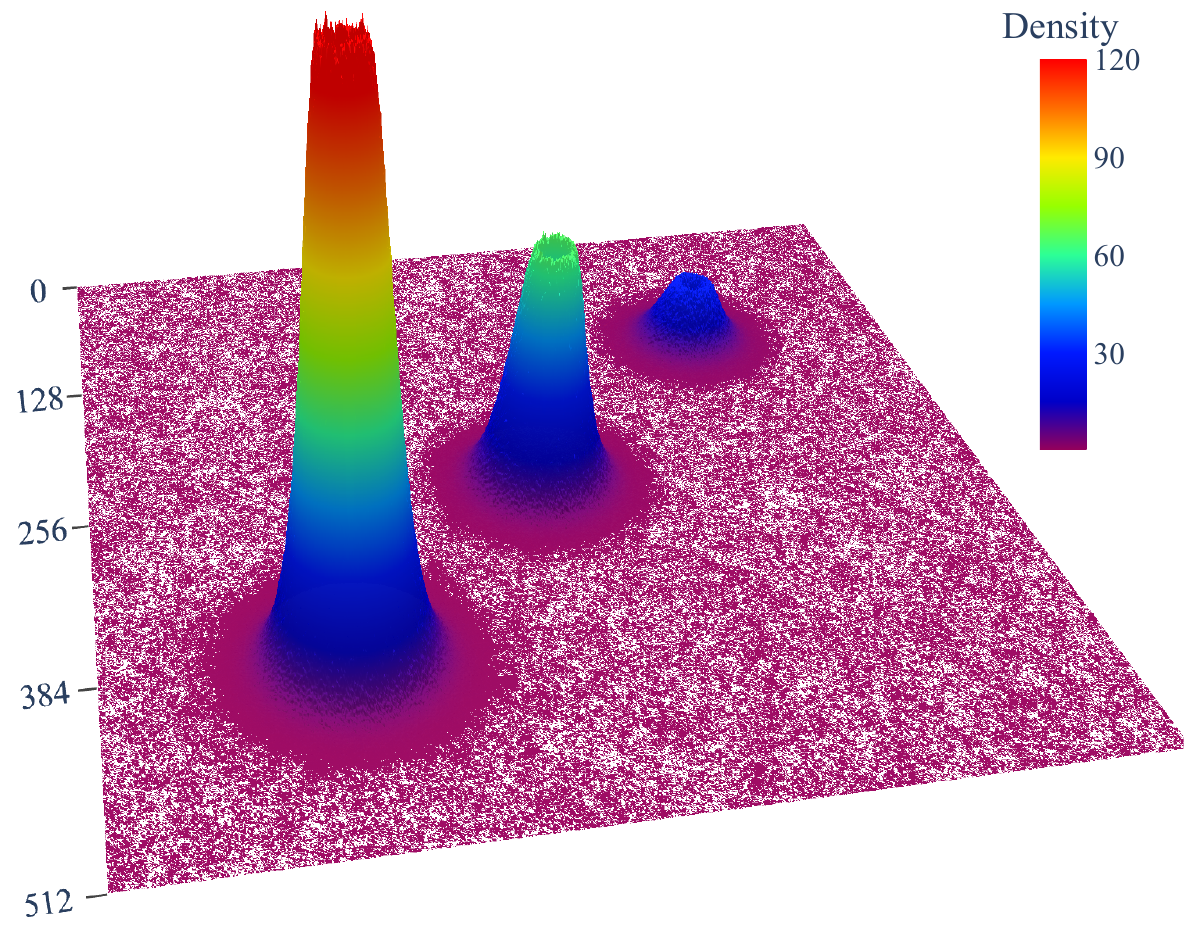}
	\caption{
	Three vortical condensates observed in systems of different sizes (unimodal ferromagnetic case
	($\omega_i=\omega_0=0.05 \;\forall i$ and  $\alpha=1$). 
	Elevation and color represent density of an average of 500 snapshots separated by 1000 time units.
	The 3 different systems ($L=256, 512, 768$) are shown 'together', taking advantage
	of the near-constant density of the surrounding gas. The peak density of the vortices is 14, 53, and 115, 
	while $\rho_0=1$. Other parameters: $D_r=0.08$, $\kappa=1$. 
	}
	\label{fig1}
\end{figure}

\begin{figure*}[t!]
\includegraphics[width=\textwidth]{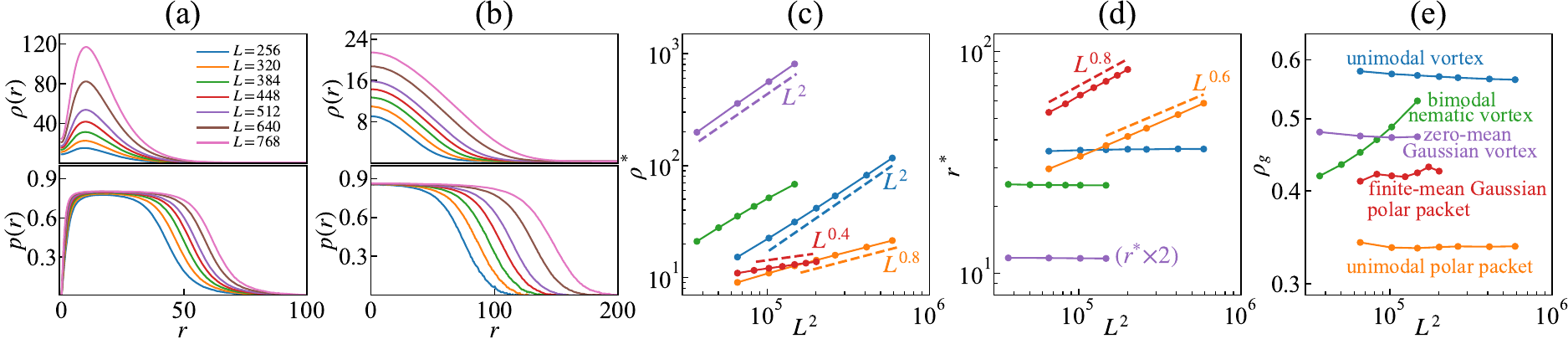}
	\caption{Condensates in systems containing a single object 
	(or two in the case of zero-mean Gaussian distribution $g(\omega)$, purple data in (c-e)). 
	All data obtained for $\rho_0=1$ and $\kappa=1$ except for bimodal nematic vortex ($\kappa=\tfrac{1}{2}$).
	All numerical details explaining how we measure quantities shown here in Appendix~B.
(a) time-averaged density (top) and order (bottom) radial profiles of polar vortices 
for the unimodal case of Fig.~\ref{fig1} ($L=256$, 320, 384, 448, 512, 640, 768).
(b) same as (a) but for rotating polar packets ($\omega_0=0.1$, $D_r=0.06$). 
(The tails of these profiles collapse when plotted vs $r/r^*$.)
(c-e): scaling of $\rho^*$, $r^*$, and $\rho_g$ with $L^2$ for 
4 different cases with ferromagnetic alignment and a nematic case: 
(i,ii) blue and orange: unimodal vortex and polar packet (same cases as in (a,b)).
(iii) red: polar packet with a Gaussian distribution $g(\omega)$ of mean 0.25 and width 0.01 ($D_r=0.07$).
(iv) purple: vortex for zero-mean Gaussian $g(\omega)$ of width 0.3 ($D_r=0.07$).
(v) green: nematic vortex ($\alpha=2$) for a bimodal distribution 
$\omega_i=\pm\omega_0$ with $\omega_0=0.06$ ($D_r=0.016$).
	Dashed lines indicate power laws with exponents close to scaling observed.
	}
	\label{fig2}
\end{figure*}

In this Letter, we show that, when made of pointwise chiral circle swimmers, 
2D dry aligning active matter does {\it not} 
follow the usual phase separation scenario.
The dense and ordered localized structures that emerge, 
which take the form of vortices or rotating polar packets,
are not microphases with a typical size. Rather, they are condensates: a single of these objects can absorb 
a finite fraction of the particles in the system, and keep a finite or slowly growing size as its mass increases.
Our results are obtained both at particle level and in continuous kinetic or hydrodynamic theories derived from
microscopic models. We consider the two main types of alignment, ferromagnetic and nematic, and both identical
and disordered chiralities, and find condensation in most cases.
With nematic alignment, however, we expect a (huge) upper limit to the size of condensates.
For polar packets, which have macroscopic order along a globally rotating axis, condensation implies phase synchronization of the particles involved. For vortices, one has only frequency synchronization. 
This occurs even with chirality disorder, and constitutes a rare case of synchronization in 2D systems 
with only local interactions. 
We then show that condensation is ultimately rooted in local-order-dependent effective diffusivity and attraction
by showing that a phenomenological theory based 
on these mechanisms accounts qualitatively for our observations. 
All information regarding the numerical methods and protocols used are in Appendix~B.

We first present results at particle level obtained using 
``Kuramoto-Vicsek" models~\footnote{These models have 
been investigated before, notably by some of us in \cite{ventejou2021susceptibility}, but also in 
\cite{liebchen2017collective,levis2019activity,levis2019simultaneous} with the only difference there that 
the interaction/alignment term was not normalized by the number of neighbors.}.
Point particles $i=1,\ldots,N$, endowed with an intrinsic frequency $\omega_i$ drawn from a 
distribution $g(\omega)$, move at constant unit speed and locally align their velocities in the presence of 
rotational noise. Their positions ${\bf r}_i$ and orientations $\theta_i$ evolve in continuous time:
\begin{align}
\dot{\bf r}_i &= {\bf e}(\theta_i) \label{eq:kvm1} \\
\dot{\theta}_i &= \omega_i + \kappa \left\langle \sin\alpha(\theta_j - \theta_i) \right\rangle_{j\sim i} + \sqrt{2D_r}\eta_i \label{eq:kvm2}
\end{align}
where ${\bf e}(\theta)$ is the unit vector along $\theta$, 
the average $\langle\ldots\rangle_{j\sim i}$ is taken over all particles within unit distance of ${\bf r}_i$, 
and $\eta_i$ is a uniform white noise drawn in $[-\pi,\pi]$.
For global all-to-all coupling, Eq.~\eqref{eq:kvm2} is a (noisy) Kuramoto model. 
In the absence of noise, a single particle describes a circle of radius $1/\omega_i$ with chirality given be the sign of $\omega_i$. In the absence of chirality ($\omega_i=0, \forall i$), one recovers a standard Vicsek model
when $\alpha=1$ (ferromagnetic alignment), or its nematic 'rods' version when $\alpha=2$. 
Below we consider both types of alignment.

Like in all Vicsek-style models, the two main parameters are the rotational noise strength $D_r$ (or, equivalently, the coupling strength $\kappa$) and the global density of particles $\rho_0=N/S$ where $S$ is the surface of the domain in which particles evolve. 
The frequency distribution $g(\omega)$ is of course also a crucial ingredient. 
Here we present results on the zero-mean distributions already considered in \cite{ventejou2021susceptibility}, namely Gaussian and bimodal ($\omega_i=\pm\omega_0$ in equal numbers).
It was shown in \cite{ventejou2021susceptibility} that (i) any finite-width $g(\omega)$
modifies qualitatively the typical phase diagram of identical achiral particles ($\omega_i=0, \, \forall i$), 
and (ii) localized structures, rotating polar packets and/or vortices, emerge at sufficiently large chiralities. 
These objects are also observed when 
considering distributions $g(\omega)$ with a finite mean, e.g. the pure
($\omega_i=\omega_0, \, \forall i$), and the disordered (Gaussian of mean $\omega_0$ and width $\delta\omega$) unimodal cases (see some cases below). 
Hereafter, we focus exclusively on these vortices and rotating polar packets.

Vortices are axisymmetric structures with either local polar or nematic order, depending on the type of 
alignment~\footnote{They were observed in all cases studied except with nematic alignment ($\alpha=2$) and finite-mean $g(\omega)$, which display more complicated objects that will be described elsewhere.}.
Polar packets, as their name suggest, are only observed with ferromagnetic alignment ($\alpha=1$); 
as reported in \cite{ventejou2021susceptibility}, they emerge at typically larger chiralities than vortices, 
and are absent in the case of a zero-mean Gaussian $g(\omega)$. 
For zero-mean but finite-width $g(\omega)$, ferromagnetic alignment induces chirality sorting, leading to, e.g.,
clockwise and counter clockwise polar vortices coexisting in space.

Simulations in large domains starting from random locations and orientations, typically produce configurations comprising many vortices or rotating polar packets, 
with their number increasing with system size, suggesting micro-phase separation. 
This is misleading: starting from a moderate size system containing a single object (or two of opposite
chiralities in the case of zero-mean $g(\omega)$ and ferromagnetic alignment), increasing progressively
the system size while maintaining the global density constant does {\it not} increase the number of structures.
Instead, the initial object(s) gets denser and denser, absorbing more and more particles, while remaining
essentially localized (Fig.~\ref{fig1}). Typical density and order profiles are shown in Fig.~\ref{fig2}(a,b).
Such condensates typically do not break into smaller structures. On the contrary, in configurations with 
many initially-formed objects, one observes, on long time scales, a decrease of the number of these objects, with dynamics
typical of a ripening process (see movies in \cite{SUPP}). The condensates are thus the most stable solutions.

In minimal configurations containing a single or a pair of opposite chirality condensates, 
the maximal density $\rho^*$ and average radius $r^*$ of these objects typically scale
 with the total number of particles $N$: 
 \begin{equation} \label{scaling}
 \rho^* \sim N^\beta \;\; {\rm and}\;\; r^* \sim N^\gamma \;\;{\rm with} \;\; \beta + 2\gamma \simeq 1\,,
 \end{equation}
which means that they contain a finite fraction of the total mass (Fig.~\ref{fig2}(c,d), data in various colors).
Vortices are strictly localized: $\beta\simeq 1$ and $\gamma\simeq 0$, with their size essentially given
by the radius of the typical `bare' circle described by an isolated particle (purple, green, and blue data).
On the other hand, the surface occupied by polar packets grows sub-extensively with $N$. In such cases, 
$0<\gamma<\tfrac{1}{2}$ and thus $0<\beta<1$. Our results indicate that these exponents are not universal 
(orange and red data).

A closer inspection of how the density $\rho_g$ of the residual gas surrounding condensates 
varies with $N$ reveals a fundamental difference between the two types of alignment (Fig.~\ref{fig2}(e)). 
For ferromagnetic alignment, $\rho_g$ converges to a finite value as $N\to\infty$. 
But for nematic vortices (green data), $\rho_g$ increases slowly with $N$.
Extrapolating this trend to larger $N$ values, $\rho_g$ must eventually reach the ordering threshold, 
i.e. the $\rho_0$ value beyond which the homogeneous disordered gas solution disappears. 
When this happens, new vortices must be nucleated.
Thus, contrary to the polarly ordered objects, nematic vortices are not true condensates, although each of them
can probably support millions of particles at the parameter values we considered.

Polar vortices and rotating packets can absorb a finite fraction of particles in the $N\to\infty$ limit. 
This is true in particular even in the presence of chirality disorder (Fig.~\ref{fig2}(c-e)). In such cases, the system is 
synchronized in the sense of the Kuramoto model, even though we are in 2D and interactions are strictly 
local~\footnote{Note, however, that an arbitrarily large number of particles can be involved locally.}. 
For rotating packets, one has phase synchronization (finite global polar $p\equiv|\langle\exp i\theta_j\rangle_j|$ when $N\to\infty$), whereas vortices only show frequency synchronization. This is a remarkable result given that
the Kuramoto model with local coupled oscillators fixed at the nodes of $d$-dimensional lattices shows phase (resp. frequency) synchronization for $d>4$ (resp. $d>2$) \cite{hong2007entrainment}.

\begin{figure}
\includegraphics[width=\columnwidth]{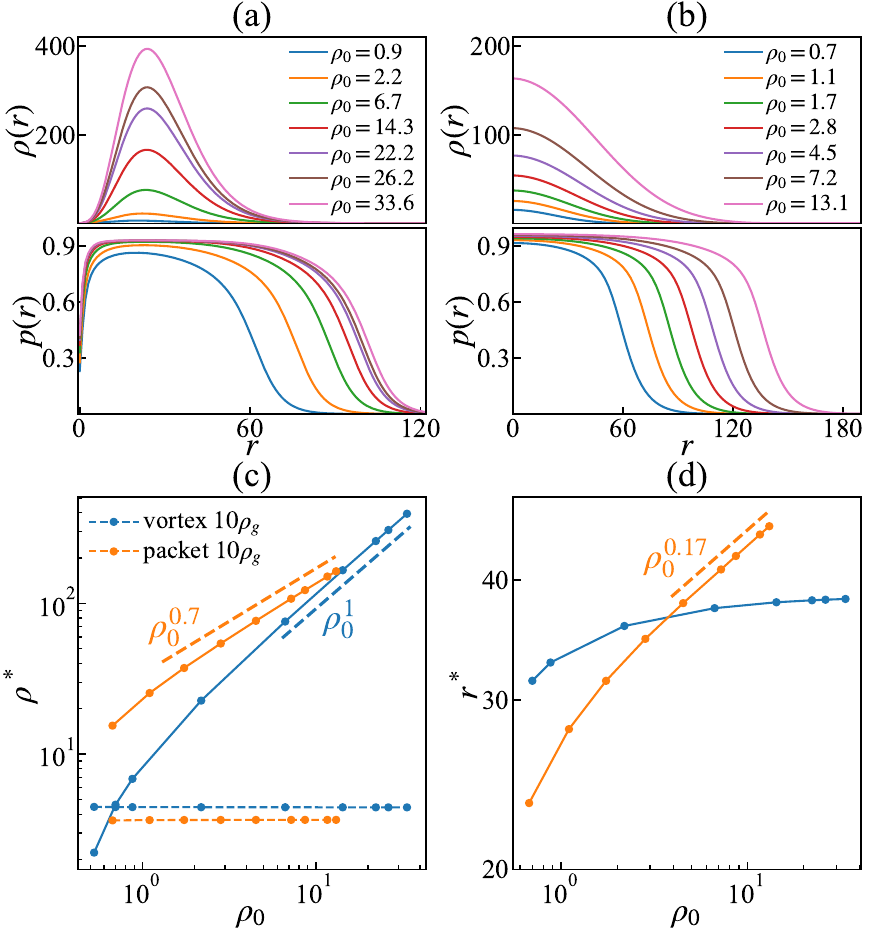}
	\caption{Vortices and polar packets observed in the truncation at order $n=5$ of the Boltzmann equation
	\eqref{eqB} (unimodal chirality $\omega_0$, ferromagnetic alignment) 
	for various $\rho_0$ values at fixed system size.
	(a) density (top) and order (bottom) radial profiles of vortex 
	($\omega_0=0.03$, $\sigma=0.5$, $D_0=0.025$, $L=256$).
	(b) same as (a) but for polar packets observed for $\omega_0=0.3$, $\sigma=0.5$, $D_0=0.05$, $L=384$.	(c,d) maximum density $\rho^*$ and $10\times\rho_g$ (c) and condensate radius $r^*$ (d) vs $\rho_0$ for data shown in (a,b).
	Dashed lines indicate power laws with exponents close to scaling observed.
	}
	\label{fig3}
\end{figure}

The above results are well accounted for by continuous kinetic and hydrodynamic descriptions
derived from the particle model.
We follow the now well-established Boltzmann-Ginzburg-Landau approach, already used in \cite{ventejou2021susceptibility} for the
case of a $\pm\omega_0$ bimodal distribution of chiralities and ferromagnetic alignment. 
Below, we only sketch the main steps followed, and focus on the simple case of a 
unimodal distribution of chiralities ($\omega_i=\omega_0, \, \forall i$).
(A more detailed exposition is given in Appendix~A.) 
The starting point is a Boltzmann equation governing the evolution of the one-body probability density function
$f({\bf r},\theta,t)$:
\begin{equation} \label{eqB}
\partial_t f + v_0{\bf e}(\theta)\cdot\nabla f +\omega_0 \partial_\theta f -D_0\nabla^2 f= 
 I_{\rm sd}[f] + I_{\rm co}[f,f],
\end{equation}
where $I_{\rm sd}$ and $I_{\rm co}$ are self-diffusion and collision integrals, the latter encoding alignment
interactions, while advection, rotation, and diffusion are on the left hand side~\footnote{An explicit spatial diffusion term with a small coefficient (typically $D_0=0.05$) has been introduced to obtain more stable results.}.
Expanding $f$ in Fourier series of $\theta$ 
(i.e. $f({\bf r},\theta,t)=\tfrac{1}{2\pi}\sum_{k=-\infty}^{+\infty} f_k({\bf r},t)e^{-ik\theta}$)
this kinetic equation is transformed into a hierarchy of partial differential equations for the complex fields $f_k$, 
with $f_0$ being nothing but the density field $\rho$, and its equation the classic exact conservation equation.
As usual with ferromagnetic alignment, only the polar momentum field $f_1$ can grow at linear level, 
and is assumed to be small at onset: $f_1\sim\epsilon$. 
This imposes the scaling ansatz $f_k\sim\epsilon^k$, $\rho-\rho_0\sim\triangledown\sim\partial_t\sim\epsilon$.
Here we do not truncate and close the hierarchy at order $\epsilon^3$ as done usually, but simply truncate 
at order $\epsilon^n$, yielding $n$ coupled nonlinear partial differential equations. This yields better, more
stable, results in numerical simulations that are essentially unchanged as soon as $n>4$. In the following we only present results obtained for $n=5$.

Simulations of the equations scanning the microscopic-level parameters 
$D_r$ and $\omega_0$ in large domains yield a phase diagram qualitatively similar to that obtained with the
original particle model (\ref{eq:kvm1},\ref{eq:kvm2}) \cite{TBP}, itself similar to that presented in \cite{ventejou2021susceptibility}
for the bimodal $\pm\omega_0$ case.
In particular, one observes vortices at small $\omega_0$ values and polar packets for stronger chiralities. 

Following the particle-level results presented above, 
we studied systematically configurations containing a single of these objects, increasing the global density $\rho_0$,
all parameters being fixed~\footnote{Similar results are obtained at fixed $\rho_0$ increasing $L$}. 
Remarkably, we also observe condensation, something never observed before with similarly-derived equations: 
our continuous equations support localized solutions 
with an apparently unbounded density peak as $\rho_0$ is increased~\footnote{As the local density gradients become large, numerical resolution has to be increased.}. 
Typical density and order
radial profiles are presented in Fig.~\ref{fig3}(a,b) for vortices and polar packets, respectively.
Similarly to our findings with the particle model, the maximal density $\rho^*$ and typical size $r^*$ 
of condensates scale with
$\rho_0$ as in Eq.~\eqref{scaling} (Fig.~\ref{fig3}(c,d)). Vortices are strictly localized ($\beta\simeq 1$, $\gamma\simeq0$), 
while polar packets are only weakly localized ($0<\gamma<\tfrac{1}{2}$).
In both cases, $\rho_g$, the density of the residual gas surrounding the condensate, 
converges to a finite value when $L\to\infty$.

A similar approach can be implemented for cases with nematic alignment. Details will be published elsewhere \cite{TBP}.
Again, we find qualitative agreement with the particle-level results. 
In particular we studied the nematic vortices arising in the case of a bimodal $\pm\omega_0$ distribution of
chiralities. As for the particle model (Fig.~\ref{fig2}), these vortices behave like strictly localized condensates, 
but the gas density $\rho_g$ increases slowly and quasi-linearly with $\rho_0$ or $L^2$. 
It is numerically relatively easy to reach the system size (or global density $\rho_0$) where $\rho_g$
passes the nematic ordering threshold. We then see the nucleation of a ring of new vortices all around the previously
isolated condensate.

Simulations of our kinetic/hydrodynamic deterministic equations thus capture the phenomena observed at particle level, indicating that fluctuations play a minor role. 
Nevertheless, such continuous descriptions do not provide much insight into the 
mechanisms giving rise to condensation.
We now propose a phenomenological theory which, we believe, fills that gap. 
For simplicity, we only consider the unimodal case, and describe our system 
averaged over the basic period, in terms of a density field $\rho$ and a scalar order field $s\in [0,1]$ 
(representing, e.g., the modulus of local polar order). 

We performed a careful analysis, both at particle- and hydrodynamic-level, 
of the fate of particles located inside condensates, where local polar order is high. 
Its details will be published elsewhere \cite{TBP},
but the main finding is a systematic drift inward (towards the condensate's center) and a strong decrease of effective diffusion compared to that measured in the gas outside.
The inward drift amounts to some effective attraction between particles in ordered regions.

\begin{figure}
\includegraphics[width=\columnwidth]{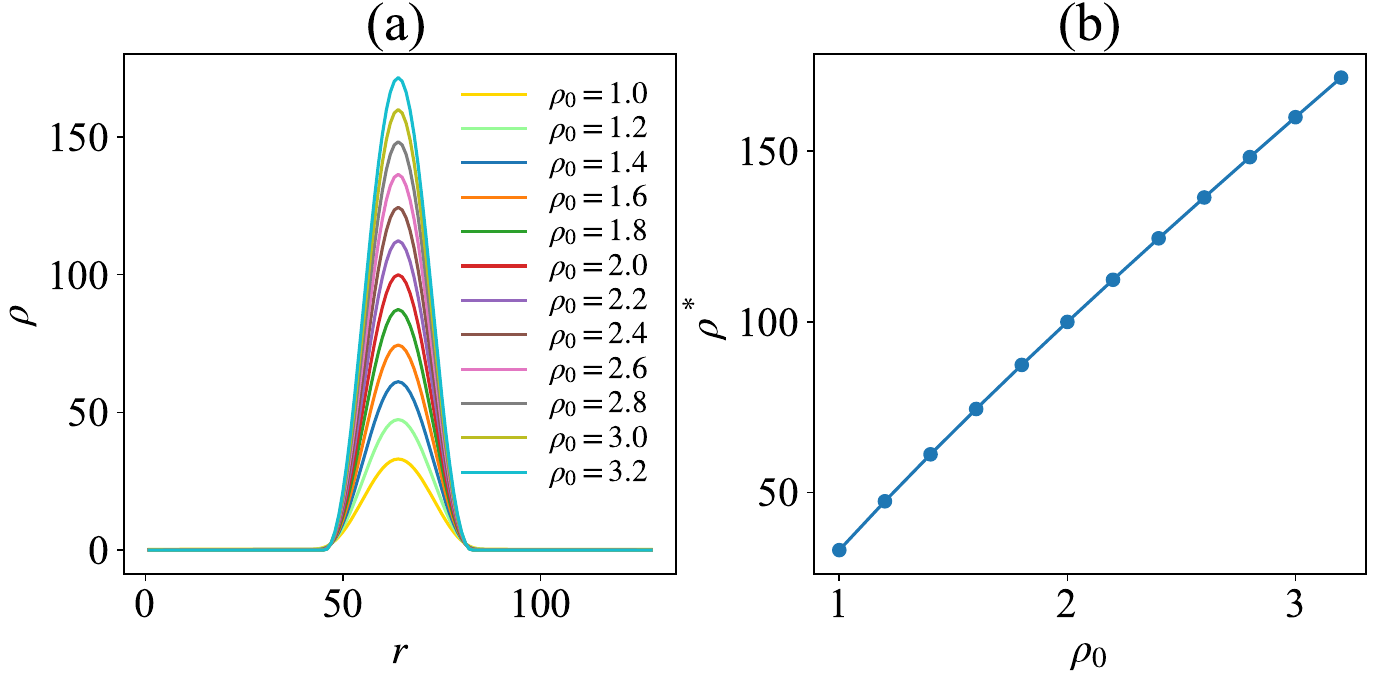}
	\caption{Condensates observed with the phenomenological model defined by Eqs.~(\ref{eq:theo1},\ref{eq:theo2}) ($D_\rho=D_s=1$, $a=0.05$, $b=1$, $c=1.25$, $L=128$).
	(a): density profiles obtained at different global densities $\rho_0$.
	(b): scaling of maximum density $\rho^*$ with $\rho_0$ for the data shown in (a).
  	}
	\label{fig4}
\end{figure}

A simple theory accounting for these observations can be written:
\begin{align}
\partial_t\rho &= D_\rho \, \nabla^2 [ (1-s) \rho ] -a \nabla\!\cdot\! [\rho \nabla(\rho s)] 
-b \nabla \!\cdot\! (\rho \nabla^3 \rho) \label{eq:theo1} \\
\partial_t s &= (\rho-1) s - c \rho s^3 + D_s \nabla^2 s   \label{eq:theo2}
\end{align}
where all coefficients are taken positive.
Eq.~\eqref{eq:theo2} has the familiar Ginzburg-Landau form present in most hydrodynamic theories 
of vectorial active matter. 
The right hand side of \eqref{eq:theo1} first displays an order-dependent diffusion term~\footnote{Note that, in \eqref{eq:theo2}, we take $c>1$, ensuring that $s<1$, so that the diffusion term in \eqref{eq:theo1} never vanishes.},
then a second term encoding attraction in ordered regions, 
while the last term accounts for surface tension effects.
This term is necessary: without attraction ($a=0$), Eqs.~(\ref{eq:theo1},\ref{eq:theo2}) yield standard phase separation for global densities $\rho_0>1$. With attraction but without surface tension ($a>0$, $b=0$), 
they lead to total, singular, condensation of all mass in one point~\footnote{Note that then our equations 
are formally similar to a Keller-Segel model potentially leading to chemotactic collapse \cite{arumugam2020keller}.}. The surface tension term corrects this unrealistic feature.

Numerical integration of Eqs.~(\ref{eq:theo1},\ref{eq:theo2}) starting from unimodal initial conditions 
yield stationary axisymmetric solutions (Fig.~\ref{fig4}). 
They are well-localized condensates: increasing global density $\rho_0$, their size $r^*$ does not increase, as shown in
Fig.~\ref{fig4}(a), 
their maximum density $\rho^*\propto\rho_0$ (Fig.~\ref{fig4}(b)), while the surrounding gas density remains approximately constant. 
This is the typical behavior of the vortices observed at particle level, 
but the density profiles shown in Fig.~\ref{fig4}(a)
do not show their central hole. 
This is because the density field in our effective theory should be seen as that of the 
centers of the elementary circles that would be described by free particles. Representing instead these virtual circles,
the density profiles do not have an empty core anymore (see examples in \cite{SUPP}).

Even though they deal with a simple situation, we believe the results shown in Fig.~\ref{fig4} allow us
to conclude that the origin of the condensation phenomena reported here lies in the reduced diffusivity
of particles in ordered regions combined with their effective attraction. 

To summarize, we have provided ample evidence that chiral, aligning, dry active matter can exhibit
condensation rather than phase separation. 
Condensation is strict with ferromagnetic alignment, and takes a weaker form
for nematic vortices. 
All condensates, be they rotating polar packets or vortices, remain extended yet synchronized collectives  
spanning many interaction lengths.
This is different from the classic zero-range process \cite{Evans_2005} and a recent active matter model in this class \cite{nava-sedeno2023vectorial} where sharp localization occurs.
This is also different from systems in which condensation arises from the presence of a strict mobility edge 
\cite{golestanian2019bose-einstein,mahault2020bose-einstein,meng2021magnetic}. In the systems treated here,
the effective diffusivity of particles decreases sharply inside condensates but does not vanish. 
In fact, the phenomena reported here can resist additional, 
explicit positional diffusion currently absent from Eq.~\eqref{eq:kvm1}.
In the presence of repulsive interactions, however, our condensates cannot accumulate arbitrarily many particles.
As long as steric effects are not dominant, though, very dense condensates are observable 
(preliminary but typical results are shown in Appendix~C).
The effects of diffusion and repulsion will be studied in detail elsewhere \cite{TBP}. 

We conclude with a few words about relevant experimental situations, which probably exist
given that condensation can be observed to some extent even with repulsion between particles. 
Growing colonies of chiral bacteria sometimes display large vortices that protrude in the third dimension 
\cite{Sirota-Madi2010genome}.
Thin chiral protein fibers driven by a carpet of molecular motors do show localized vortices \cite{denk2016active}.
We believe that in both cases, some condensation effects might be at play. 
It could also be interesting to see whether magnetic microswimmers performing chiral motion can condensate 
(some experiments in external field have shown promising prospects \cite{meng2021magnetic}). 
We look forward to seeing some of these systems explored accordingly.

\acknowledgments
We thank Yu Duan, Beno\^{\i}t Mahault, and Yongfeng Zhao for useful remarks. 
We acknowledge generous allocations of cpu time on Beijing CSRC’s Tianhe supercomputer.
This work is supported by the National Natural Science Foundation of China (Grants No. 12275188 and No. 11922506).

\bibliography{./../Biblio-current.bib}

\begin{thebibliography}{44}%
\makeatletter
\providecommand \@ifxundefined [1]{%
 \@ifx{#1\undefined}
}%
\providecommand \@ifnum [1]{%
 \ifnum #1\expandafter \@firstoftwo
 \else \expandafter \@secondoftwo
 \fi
}%
\providecommand \@ifx [1]{%
 \ifx #1\expandafter \@firstoftwo
 \else \expandafter \@secondoftwo
 \fi
}%
\providecommand \natexlab [1]{#1}%
\providecommand \enquote  [1]{``#1''}%
\providecommand \bibnamefont  [1]{#1}%
\providecommand \bibfnamefont [1]{#1}%
\providecommand \citenamefont [1]{#1}%
\providecommand \href@noop [0]{\@secondoftwo}%
\providecommand \href [0]{\begingroup \@sanitize@url \@href}%
\providecommand \@href[1]{\@@startlink{#1}\@@href}%
\providecommand \@@href[1]{\endgroup#1\@@endlink}%
\providecommand \@sanitize@url [0]{\catcode `\\12\catcode `\$12\catcode
  `\&12\catcode `\#12\catcode `\^12\catcode `\_12\catcode `\%12\relax}%
\providecommand \@@startlink[1]{}%
\providecommand \@@endlink[0]{}%
\providecommand \url  [0]{\begingroup\@sanitize@url \@url }%
\providecommand \@url [1]{\endgroup\@href {#1}{\urlprefix }}%
\providecommand \urlprefix  [0]{URL }%
\providecommand \Eprint [0]{\href }%
\providecommand \doibase [0]{http://dx.doi.org/}%
\providecommand \selectlanguage [0]{\@gobble}%
\providecommand \bibinfo  [0]{\@secondoftwo}%
\providecommand \bibfield  [0]{\@secondoftwo}%
\providecommand \translation [1]{[#1]}%
\providecommand \BibitemOpen [0]{}%
\providecommand \bibitemStop [0]{}%
\providecommand \bibitemNoStop [0]{.\EOS\space}%
\providecommand \EOS [0]{\spacefactor3000\relax}%
\providecommand \BibitemShut  [1]{\csname bibitem#1\endcsname}%
\let\auto@bib@innerbib\@empty
\bibitem [{\citenamefont {Cates}\ and\ \citenamefont
  {Tailleur}(2015)}]{cates2015motility}%
  \BibitemOpen
  \bibfield  {author} {\bibinfo {author} {\bibfnamefont {M.~E.}\ \bibnamefont
  {Cates}}\ and\ \bibinfo {author} {\bibfnamefont {J.}~\bibnamefont
  {Tailleur}},\ }\href
  {http://www.annualreviews.org/doi/abs/10.1146/annurev-conmatphys-031214-014710}
  {\bibfield  {journal} {\bibinfo  {journal} {Annu. Rev. Condens. Matter
  Phys.}\ }\textbf {\bibinfo {volume} {6}},\ \bibinfo {pages} {219} (\bibinfo
  {year} {2015})}\BibitemShut {NoStop}%
\bibitem [{\citenamefont {Chat{\'e}}(2020)}]{chate2020dry}%
  \BibitemOpen
  \bibfield  {author} {\bibinfo {author} {\bibfnamefont {H.}~\bibnamefont
  {Chat{\'e}}},\ }\href {\doibase 10.1146/annurev-conmatphys-031119-050752}
  {\bibfield  {journal} {\bibinfo  {journal} {Annu. Rev. Cond. Matt.}\ }\textbf
  {\bibinfo {volume} {11}},\ \bibinfo {pages} {189} (\bibinfo {year}
  {2020})}\BibitemShut {NoStop}%
\bibitem [{Note1()}]{Note1}%
  \BibitemOpen
  \bibinfo {note} {In systems with long-range interactions, however, such as
  those governed by chemotaxis, condensation can occur (chemotactic collapse).
  See, e.g., \cite {saha2014clusters,pohl2014dynamic}}\BibitemShut {NoStop}%
\bibitem [{Note2()}]{Note2}%
  \BibitemOpen
  \bibinfo {note} {See \cite {Liebchen_2022} for a short recent review and
  references therein.}\BibitemShut {Stop}%
\bibitem [{\citenamefont {Schwarz-Linek}\ \emph {et~al.}(2012)\citenamefont
  {Schwarz-Linek}, \citenamefont {Valeriani}, \citenamefont {Cacciuto},
  \citenamefont {Cates}, \citenamefont {Marenduzzo}, \citenamefont {Morozov},\
  and\ \citenamefont {Poon}}]{schwarz2012phase}%
  \BibitemOpen
  \bibfield  {author} {\bibinfo {author} {\bibfnamefont {J.}~\bibnamefont
  {Schwarz-Linek}}, \bibinfo {author} {\bibfnamefont {C.}~\bibnamefont
  {Valeriani}}, \bibinfo {author} {\bibfnamefont {A.}~\bibnamefont {Cacciuto}},
  \bibinfo {author} {\bibfnamefont {M.}~\bibnamefont {Cates}}, \bibinfo
  {author} {\bibfnamefont {D.}~\bibnamefont {Marenduzzo}}, \bibinfo {author}
  {\bibfnamefont {A.}~\bibnamefont {Morozov}}, \ and\ \bibinfo {author}
  {\bibfnamefont {W.}~\bibnamefont {Poon}},\ }\href@noop {} {\bibfield
  {journal} {\bibinfo  {journal} {Proceedings of the National Academy of
  Sciences}\ }\textbf {\bibinfo {volume} {109}},\ \bibinfo {pages} {4052}
  (\bibinfo {year} {2012})}\BibitemShut {NoStop}%
\bibitem [{\citenamefont {Denk}\ \emph {et~al.}(2016)\citenamefont {Denk},
  \citenamefont {Huber}, \citenamefont {Reithmann},\ and\ \citenamefont
  {Frey}}]{denk2016active}%
  \BibitemOpen
  \bibfield  {author} {\bibinfo {author} {\bibfnamefont {J.}~\bibnamefont
  {Denk}}, \bibinfo {author} {\bibfnamefont {L.}~\bibnamefont {Huber}},
  \bibinfo {author} {\bibfnamefont {E.}~\bibnamefont {Reithmann}}, \ and\
  \bibinfo {author} {\bibfnamefont {E.}~\bibnamefont {Frey}},\ }\href
  {https://journals.aps.org/prl/abstract/10.1103/PhysRevLett.116.178301}
  {\bibfield  {journal} {\bibinfo  {journal} {Physical Review Letters}\
  }\textbf {\bibinfo {volume} {116}},\ \bibinfo {pages} {178301} (\bibinfo
  {year} {2016})}\BibitemShut {NoStop}%
\bibitem [{\citenamefont {Liebchen}\ \emph {et~al.}(2016)\citenamefont
  {Liebchen}, \citenamefont {Cates},\ and\ \citenamefont
  {Marenduzzo}}]{liebchen2016pattern}%
  \BibitemOpen
  \bibfield  {author} {\bibinfo {author} {\bibfnamefont {B.}~\bibnamefont
  {Liebchen}}, \bibinfo {author} {\bibfnamefont {M.~E.}\ \bibnamefont {Cates}},
  \ and\ \bibinfo {author} {\bibfnamefont {D.}~\bibnamefont {Marenduzzo}},\
  }\href {http://pubs.rsc.org/-/content/articlehtml/2016/sm/c6sm01162d}
  {\bibfield  {journal} {\bibinfo  {journal} {Soft Matter}\ }\textbf {\bibinfo
  {volume} {12}},\ \bibinfo {pages} {7259} (\bibinfo {year}
  {2016})}\BibitemShut {NoStop}%
\bibitem [{\citenamefont {{Huang}}\ \emph {et~al.}(2020)\citenamefont
  {{Huang}}, \citenamefont {{Menzel}},\ and\ \citenamefont
  {{L{\"o}wen}}}]{huang2020dynamical}%
  \BibitemOpen
  \bibfield  {author} {\bibinfo {author} {\bibfnamefont {Z.-F.}\ \bibnamefont
  {{Huang}}}, \bibinfo {author} {\bibfnamefont {A.~M.}\ \bibnamefont
  {{Menzel}}}, \ and\ \bibinfo {author} {\bibfnamefont {H.}~\bibnamefont
  {{L{\"o}wen}}},\ }\href {\doibase 10.1103/PhysRevLett.125.218002} {\bibfield
  {journal} {\bibinfo  {journal} {\prl}\ }\textbf {\bibinfo {volume} {125}},\
  \bibinfo {eid} {218002} (\bibinfo {year} {2020})},\ \Eprint
  {http://arxiv.org/abs/2010.01252} {arXiv:2010.01252 [cond-mat.soft]}
  \BibitemShut {NoStop}%
\bibitem [{\citenamefont {Ventejou}\ \emph {et~al.}(2021)\citenamefont
  {Ventejou}, \citenamefont {Chat\'e}, \citenamefont {Montagne},\ and\
  \citenamefont {Shi}}]{ventejou2021susceptibility}%
  \BibitemOpen
  \bibfield  {author} {\bibinfo {author} {\bibfnamefont {B.}~\bibnamefont
  {Ventejou}}, \bibinfo {author} {\bibfnamefont {H.}~\bibnamefont {Chat\'e}},
  \bibinfo {author} {\bibfnamefont {R.}~\bibnamefont {Montagne}}, \ and\
  \bibinfo {author} {\bibfnamefont {X.-q.}\ \bibnamefont {Shi}},\ }\href
  {\doibase 10.1103/PhysRevLett.127.238001} {\bibfield  {journal} {\bibinfo
  {journal} {Phys. Rev. Lett.}\ }\textbf {\bibinfo {volume} {127}},\ \bibinfo
  {pages} {238001} (\bibinfo {year} {2021})}\BibitemShut {NoStop}%
\bibitem [{\citenamefont {Moore}\ \emph {et~al.}(2021)\citenamefont {Moore},
  \citenamefont {Glaser},\ and\ \citenamefont {Betterton}}]{moore2021chiral}%
  \BibitemOpen
  \bibfield  {author} {\bibinfo {author} {\bibfnamefont {J.~M.}\ \bibnamefont
  {Moore}}, \bibinfo {author} {\bibfnamefont {M.~A.}\ \bibnamefont {Glaser}}, \
  and\ \bibinfo {author} {\bibfnamefont {M.~D.}\ \bibnamefont {Betterton}},\
  }\href {\doibase 10.1039/D0SM01163K} {\bibfield  {journal} {\bibinfo
  {journal} {Soft Matter}\ }\textbf {\bibinfo {volume} {17}},\ \bibinfo {pages}
  {4559} (\bibinfo {year} {2021})}\BibitemShut {NoStop}%
\bibitem [{\citenamefont {Scholz}\ \emph {et~al.}(2021)\citenamefont {Scholz},
  \citenamefont {Ldov}, \citenamefont {P{\"o}schel}, \citenamefont {Engel},\
  and\ \citenamefont {L{\"o}wen}}]{scholz2021surfactants}%
  \BibitemOpen
  \bibfield  {author} {\bibinfo {author} {\bibfnamefont {C.}~\bibnamefont
  {Scholz}}, \bibinfo {author} {\bibfnamefont {A.}~\bibnamefont {Ldov}},
  \bibinfo {author} {\bibfnamefont {T.}~\bibnamefont {P{\"o}schel}}, \bibinfo
  {author} {\bibfnamefont {M.}~\bibnamefont {Engel}}, \ and\ \bibinfo {author}
  {\bibfnamefont {H.}~\bibnamefont {L{\"o}wen}},\ }\href@noop {} {\bibfield
  {journal} {\bibinfo  {journal} {Science Advances}\ }\textbf {\bibinfo
  {volume} {7}},\ \bibinfo {pages} {eabf8998} (\bibinfo {year}
  {2021})}\BibitemShut {NoStop}%
\bibitem [{\citenamefont {Massana-Cid}\ \emph {et~al.}(2021)\citenamefont
  {Massana-Cid}, \citenamefont {Levis}, \citenamefont {Hern\'andez},
  \citenamefont {Pagonabarraga},\ and\ \citenamefont
  {Tierno}}]{massana-cid2021arrested}%
  \BibitemOpen
  \bibfield  {author} {\bibinfo {author} {\bibfnamefont {H.}~\bibnamefont
  {Massana-Cid}}, \bibinfo {author} {\bibfnamefont {D.}~\bibnamefont {Levis}},
  \bibinfo {author} {\bibfnamefont {R.~J.~H.}\ \bibnamefont {Hern\'andez}},
  \bibinfo {author} {\bibfnamefont {I.}~\bibnamefont {Pagonabarraga}}, \ and\
  \bibinfo {author} {\bibfnamefont {P.}~\bibnamefont {Tierno}},\ }\href
  {\doibase 10.1103/PhysRevResearch.3.L042021} {\bibfield  {journal} {\bibinfo
  {journal} {Phys. Rev. Res.}\ }\textbf {\bibinfo {volume} {3}},\ \bibinfo
  {pages} {L042021} (\bibinfo {year} {2021})}\BibitemShut {NoStop}%
\bibitem [{\citenamefont {Hokmabad}\ \emph {et~al.}(2022)\citenamefont
  {Hokmabad}, \citenamefont {Nishide}, \citenamefont {Ramesh}, \citenamefont
  {Kr{\"u}ger},\ and\ \citenamefont {Maass}}]{hokmabad2022spontaneously}%
  \BibitemOpen
  \bibfield  {author} {\bibinfo {author} {\bibfnamefont {B.~V.}\ \bibnamefont
  {Hokmabad}}, \bibinfo {author} {\bibfnamefont {A.}~\bibnamefont {Nishide}},
  \bibinfo {author} {\bibfnamefont {P.}~\bibnamefont {Ramesh}}, \bibinfo
  {author} {\bibfnamefont {C.}~\bibnamefont {Kr{\"u}ger}}, \ and\ \bibinfo
  {author} {\bibfnamefont {C.~C.}\ \bibnamefont {Maass}},\ }\href {\doibase
  10.1039/D1SM01795K} {\bibfield  {journal} {\bibinfo  {journal} {Soft Matter}\
  }\textbf {\bibinfo {volume} {18}},\ \bibinfo {pages} {2731} (\bibinfo {year}
  {2022})}\BibitemShut {NoStop}%
\bibitem [{\citenamefont {Ma}\ and\ \citenamefont
  {Ni}(2022)}]{ma2022dynamical}%
  \BibitemOpen
  \bibfield  {author} {\bibinfo {author} {\bibfnamefont {Z.}~\bibnamefont
  {Ma}}\ and\ \bibinfo {author} {\bibfnamefont {R.}~\bibnamefont {Ni}},\
  }\href@noop {} {\bibfield  {journal} {\bibinfo  {journal} {The Journal of
  Chemical Physics}\ }\textbf {\bibinfo {volume} {156}},\ \bibinfo {pages}
  {021102} (\bibinfo {year} {2022})}\BibitemShut {NoStop}%
\bibitem [{\citenamefont {Levis}\ \emph {et~al.}(2019)\citenamefont {Levis},
  \citenamefont {Pagonabarraga},\ and\ \citenamefont
  {Liebchen}}]{levis2019activity}%
  \BibitemOpen
  \bibfield  {author} {\bibinfo {author} {\bibfnamefont {D.}~\bibnamefont
  {Levis}}, \bibinfo {author} {\bibfnamefont {I.}~\bibnamefont
  {Pagonabarraga}}, \ and\ \bibinfo {author} {\bibfnamefont {B.}~\bibnamefont
  {Liebchen}},\ }\href {\doibase 10.1103/PhysRevResearch.1.023026} {\bibfield
  {journal} {\bibinfo  {journal} {Phys. Rev. Research}\ }\textbf {\bibinfo
  {volume} {1}},\ \bibinfo {pages} {023026} (\bibinfo {year}
  {2019})}\BibitemShut {NoStop}%
\bibitem [{\citenamefont {Kruk}\ \emph {et~al.}(2020)\citenamefont {Kruk},
  \citenamefont {Carrillo},\ and\ \citenamefont {Koeppl}}]{kruk2020traveling}%
  \BibitemOpen
  \bibfield  {author} {\bibinfo {author} {\bibfnamefont {N.}~\bibnamefont
  {Kruk}}, \bibinfo {author} {\bibfnamefont {J.~A.}\ \bibnamefont {Carrillo}},
  \ and\ \bibinfo {author} {\bibfnamefont {H.}~\bibnamefont {Koeppl}},\ }\href
  {\doibase 10.1103/PhysRevE.102.022604} {\bibfield  {journal} {\bibinfo
  {journal} {Phys. Rev. E}\ }\textbf {\bibinfo {volume} {102}},\ \bibinfo
  {pages} {022604} (\bibinfo {year} {2020})}\BibitemShut {NoStop}%
\bibitem [{\citenamefont {{Tan}}\ \emph {et~al.}(2022)\citenamefont {{Tan}},
  \citenamefont {{Mietke}}, \citenamefont {{Li}}, \citenamefont {{Chen}},
  \citenamefont {{Higinbotham}}, \citenamefont {{Foster}}, \citenamefont
  {{Gokhale}}, \citenamefont {{Dunkel}},\ and\ \citenamefont
  {{Fakhri}}}]{tan2022odd}%
  \BibitemOpen
  \bibfield  {author} {\bibinfo {author} {\bibfnamefont {T.~H.}\ \bibnamefont
  {{Tan}}}, \bibinfo {author} {\bibfnamefont {A.}~\bibnamefont {{Mietke}}},
  \bibinfo {author} {\bibfnamefont {J.}~\bibnamefont {{Li}}}, \bibinfo {author}
  {\bibfnamefont {Y.}~\bibnamefont {{Chen}}}, \bibinfo {author} {\bibfnamefont
  {H.}~\bibnamefont {{Higinbotham}}}, \bibinfo {author} {\bibfnamefont {P.~J.}\
  \bibnamefont {{Foster}}}, \bibinfo {author} {\bibfnamefont {S.}~\bibnamefont
  {{Gokhale}}}, \bibinfo {author} {\bibfnamefont {J.}~\bibnamefont {{Dunkel}}},
  \ and\ \bibinfo {author} {\bibfnamefont {N.}~\bibnamefont {{Fakhri}}},\
  }\href {\doibase 10.1038/s41586-022-04889-6} {\bibfield  {journal} {\bibinfo
  {journal} {\nat}\ }\textbf {\bibinfo {volume} {607}},\ \bibinfo {pages} {287}
  (\bibinfo {year} {2022})},\ \Eprint {http://arxiv.org/abs/2105.07507}
  {arXiv:2105.07507 [cond-mat.soft]} \BibitemShut {NoStop}%
\bibitem [{\citenamefont {Caporusso}\ \emph {et~al.}(2024)\citenamefont
  {Caporusso}, \citenamefont {Gonnella},\ and\ \citenamefont
  {Levis}}]{caporusso2024phase}%
  \BibitemOpen
  \bibfield  {author} {\bibinfo {author} {\bibfnamefont {C.~B.}\ \bibnamefont
  {Caporusso}}, \bibinfo {author} {\bibfnamefont {G.}~\bibnamefont {Gonnella}},
  \ and\ \bibinfo {author} {\bibfnamefont {D.}~\bibnamefont {Levis}},\ }\href
  {\doibase 10.1103/PhysRevLett.132.168201} {\bibfield  {journal} {\bibinfo
  {journal} {Phys. Rev. Lett.}\ }\textbf {\bibinfo {volume} {132}},\ \bibinfo
  {pages} {168201} (\bibinfo {year} {2024})}\BibitemShut {NoStop}%
\bibitem [{\citenamefont {Liebchen}\ and\ \citenamefont
  {Levis}(2017)}]{liebchen2017collective}%
  \BibitemOpen
  \bibfield  {author} {\bibinfo {author} {\bibfnamefont {B.}~\bibnamefont
  {Liebchen}}\ and\ \bibinfo {author} {\bibfnamefont {D.}~\bibnamefont
  {Levis}},\ }\href
  {https://journals.aps.org/prl/abstract/10.1103/PhysRevLett.119.058002}
  {\bibfield  {journal} {\bibinfo  {journal} {Physical Review Letters}\
  }\textbf {\bibinfo {volume} {119}},\ \bibinfo {pages} {058002} (\bibinfo
  {year} {2017})}\BibitemShut {NoStop}%
\bibitem [{\citenamefont {Levis}\ and\ \citenamefont
  {Liebchen}(2019)}]{levis2019simultaneous}%
  \BibitemOpen
  \bibfield  {author} {\bibinfo {author} {\bibfnamefont {D.}~\bibnamefont
  {Levis}}\ and\ \bibinfo {author} {\bibfnamefont {B.}~\bibnamefont
  {Liebchen}},\ }\href {\doibase 10.1103/PhysRevE.100.012406} {\bibfield
  {journal} {\bibinfo  {journal} {Phys. Rev. E}\ }\textbf {\bibinfo {volume}
  {100}},\ \bibinfo {pages} {012406} (\bibinfo {year} {2019})}\BibitemShut
  {NoStop}%
\bibitem [{\citenamefont {Kreienkamp}\ and\ \citenamefont
  {Klapp}(2022)}]{Kreienkamp_2022}%
  \BibitemOpen
  \bibfield  {author} {\bibinfo {author} {\bibfnamefont {K.~L.}\ \bibnamefont
  {Kreienkamp}}\ and\ \bibinfo {author} {\bibfnamefont {S.~H.~L.}\ \bibnamefont
  {Klapp}},\ }\href {\doibase 10.1088/1367-2630/ac9cc3} {\bibfield  {journal}
  {\bibinfo  {journal} {New Journal of Physics}\ }\textbf {\bibinfo {volume}
  {24}},\ \bibinfo {pages} {123009} (\bibinfo {year} {2022})}\BibitemShut
  {NoStop}%
\bibitem [{\citenamefont {Ai}\ \emph {et~al.}(2023)\citenamefont {Ai},
  \citenamefont {Quan},\ and\ \citenamefont {Li}}]{ai2023spontaneous}%
  \BibitemOpen
  \bibfield  {author} {\bibinfo {author} {\bibfnamefont {B.-Q.}\ \bibnamefont
  {Ai}}, \bibinfo {author} {\bibfnamefont {S.}~\bibnamefont {Quan}}, \ and\
  \bibinfo {author} {\bibfnamefont {F.-g.}\ \bibnamefont {Li}},\ }\href@noop {}
  {\bibfield  {journal} {\bibinfo  {journal} {New Journal of Physics}\ }\textbf
  {\bibinfo {volume} {25}},\ \bibinfo {pages} {063025} (\bibinfo {year}
  {2023})}\BibitemShut {NoStop}%
\bibitem [{\citenamefont {Semwal}\ \emph {et~al.}(2024)\citenamefont {Semwal},
  \citenamefont {Joshi}, \citenamefont {Dikshit},\ and\ \citenamefont
  {Mishra}}]{semwal2024macro}%
  \BibitemOpen
  \bibfield  {author} {\bibinfo {author} {\bibfnamefont {V.}~\bibnamefont
  {Semwal}}, \bibinfo {author} {\bibfnamefont {J.}~\bibnamefont {Joshi}},
  \bibinfo {author} {\bibfnamefont {S.}~\bibnamefont {Dikshit}}, \ and\
  \bibinfo {author} {\bibfnamefont {S.}~\bibnamefont {Mishra}},\ }\href
  {\doibase https://doi.org/10.1016/j.physa.2023.129435} {\bibfield  {journal}
  {\bibinfo  {journal} {Physica A: Statistical Mechanics and its Applications}\
  }\textbf {\bibinfo {volume} {634}},\ \bibinfo {pages} {129435} (\bibinfo
  {year} {2024})}\BibitemShut {NoStop}%
\bibitem [{Note3()}]{Note3}%
  \BibitemOpen
  \bibinfo {note} {These models have been investigated before, notably by some
  of us in \cite {ventejou2021susceptibility}, but also in \cite
  {liebchen2017collective,levis2019activity,levis2019simultaneous} with the
  only difference there that the interaction/alignment term was not normalized
  by the number of neighbors.}\BibitemShut {Stop}%
\bibitem [{Note4()}]{Note4}%
  \BibitemOpen
  \bibinfo {note} {They were observed in all cases studied except with nematic
  alignment ($\alpha =2$) and finite-mean $g(\omega )$, which display more
  complicated objects that will be described elsewhere.}\BibitemShut {Stop}%
\bibitem [{SUP()}]{SUPP}%
  \BibitemOpen
  \href@noop {} {\bibinfo  {journal} {{See Supplementary Material at
  [to-be-inserted-by-publisher] for...}}\ }\BibitemShut {NoStop}%
\bibitem [{Note5()}]{Note5}%
  \BibitemOpen
\bibfield  {journal} {  }\bibinfo {note} {Note, however, that an arbitrarily
  large number of particles can be involved locally.}\BibitemShut {Stop}%
\bibitem [{\citenamefont {Hong}\ \emph {et~al.}(2007)\citenamefont {Hong},
  \citenamefont {Chat\'e}, \citenamefont {Park},\ and\ \citenamefont
  {Tang}}]{hong2007entrainment}%
  \BibitemOpen
  \bibfield  {author} {\bibinfo {author} {\bibfnamefont {H.}~\bibnamefont
  {Hong}}, \bibinfo {author} {\bibfnamefont {H.}~\bibnamefont {Chat\'e}},
  \bibinfo {author} {\bibfnamefont {H.}~\bibnamefont {Park}}, \ and\ \bibinfo
  {author} {\bibfnamefont {L.-H.}\ \bibnamefont {Tang}},\ }\href {\doibase
  10.1103/PhysRevLett.99.184101} {\bibfield  {journal} {\bibinfo  {journal}
  {Phys. Rev. Lett.}\ }\textbf {\bibinfo {volume} {99}},\ \bibinfo {pages}
  {184101} (\bibinfo {year} {2007})}\BibitemShut {NoStop}%
\bibitem [{Note6()}]{Note6}%
  \BibitemOpen
  \bibinfo {note} {An explicit spatial diffusion term with a small coefficient
  (typically $D_0=0.05$) has been introduced to obtain more stable
  results.}\BibitemShut {Stop}%
\bibitem [{\citenamefont {Wang}\ \emph {et~al.}(2024)\citenamefont {Wang},
  \citenamefont {Ventejou}, \citenamefont {Chat{\'e}}, \citenamefont
  {Montagne},\ and\ \citenamefont {Shi}}]{TBP}%
  \BibitemOpen
  \bibfield  {author} {\bibinfo {author} {\bibfnamefont {Y.}~\bibnamefont
  {Wang}}, \bibinfo {author} {\bibfnamefont {B.}~\bibnamefont {Ventejou}},
  \bibinfo {author} {\bibfnamefont {H.}~\bibnamefont {Chat{\'e}}}, \bibinfo
  {author} {\bibfnamefont {R.}~\bibnamefont {Montagne}}, \ and\ \bibinfo
  {author} {\bibfnamefont {X.-q.}\ \bibnamefont {Shi}},\ }\href@noop {}
  {\enquote {\bibinfo {title} {to be published},}\ } (\bibinfo {year}
  {2024})\BibitemShut {NoStop}%
\bibitem [{Note7()}]{Note7}%
  \BibitemOpen
  \bibinfo {note} {Similar results are obtained at fixed $\rho _0$ increasing
  $L$}\BibitemShut {NoStop}%
\bibitem [{Note8()}]{Note8}%
  \BibitemOpen
  \bibinfo {note} {As the local density gradients become large, numerical
  resolution has to be increased.}\BibitemShut {Stop}%
\bibitem [{Note9()}]{Note9}%
  \BibitemOpen
  \bibinfo {note} {Note that, in \protect \textup {\hbox {\mathsurround \z@
  \protect \normalfont (\ignorespaces \ref {eq:theo2}\unskip \@@italiccorr )}},
  we take $c>1$, ensuring that $s<1$, so that the diffusion term in \protect
  \textup {\hbox {\mathsurround \z@ \protect \normalfont (\ignorespaces \ref
  {eq:theo1}\unskip \@@italiccorr )}} never vanishes.}\BibitemShut {Stop}%
\bibitem [{Note10()}]{Note10}%
  \BibitemOpen
  \bibinfo {note} {Note that then our equations are formally similar to a
  Keller-Segel model potentially leading to chemotactic collapse \cite
  {arumugam2020keller}.}\BibitemShut {Stop}%
\bibitem [{\citenamefont {Evans}\ and\ \citenamefont
  {Hanney}(2005)}]{Evans_2005}%
  \BibitemOpen
  \bibfield  {author} {\bibinfo {author} {\bibfnamefont {M.~R.}\ \bibnamefont
  {Evans}}\ and\ \bibinfo {author} {\bibfnamefont {T.}~\bibnamefont {Hanney}},\
  }\href {\doibase 10.1088/0305-4470/38/19/R01} {\bibfield  {journal} {\bibinfo
   {journal} {Journal of Physics A: Mathematical and General}\ }\textbf
  {\bibinfo {volume} {38}},\ \bibinfo {pages} {R195} (\bibinfo {year}
  {2005})}\BibitemShut {NoStop}%
\bibitem [{\citenamefont {Nava-Sede{\~n}o}\ \emph {et~al.}(2023)\citenamefont
  {Nava-Sede{\~n}o}, \citenamefont {Hatzikirou}, \citenamefont
  {Vo{\ss}-B{\"o}hme}, \citenamefont {Brusch}, \citenamefont {Deutsch},\ and\
  \citenamefont {Peruani}}]{nava-sedeno2023vectorial}%
  \BibitemOpen
  \bibfield  {author} {\bibinfo {author} {\bibfnamefont {J.~M.}\ \bibnamefont
  {Nava-Sede{\~n}o}}, \bibinfo {author} {\bibfnamefont {H.}~\bibnamefont
  {Hatzikirou}}, \bibinfo {author} {\bibfnamefont {A.}~\bibnamefont
  {Vo{\ss}-B{\"o}hme}}, \bibinfo {author} {\bibfnamefont {L.}~\bibnamefont
  {Brusch}}, \bibinfo {author} {\bibfnamefont {A.}~\bibnamefont {Deutsch}}, \
  and\ \bibinfo {author} {\bibfnamefont {F.}~\bibnamefont {Peruani}},\ }\href
  {\doibase 10.1088/1367-2630/ad1498} {\bibfield  {journal} {\bibinfo
  {journal} {New Journal of Physics}\ }\textbf {\bibinfo {volume} {25}},\
  \bibinfo {pages} {123046} (\bibinfo {year} {2023})}\BibitemShut {NoStop}%
\bibitem [{\citenamefont {Golestanian}(2019)}]{golestanian2019bose-einstein}%
  \BibitemOpen
  \bibfield  {author} {\bibinfo {author} {\bibfnamefont {R.}~\bibnamefont
  {Golestanian}},\ }\href {\doibase 10.1103/PhysRevE.100.010601} {\bibfield
  {journal} {\bibinfo  {journal} {Phys. Rev. E}\ }\textbf {\bibinfo {volume}
  {100}},\ \bibinfo {pages} {010601} (\bibinfo {year} {2019})}\BibitemShut
  {NoStop}%
\bibitem [{\citenamefont {Mahault}\ and\ \citenamefont
  {Golestanian}(2020)}]{mahault2020bose-einstein}%
  \BibitemOpen
  \bibfield  {author} {\bibinfo {author} {\bibfnamefont {B.}~\bibnamefont
  {Mahault}}\ and\ \bibinfo {author} {\bibfnamefont {R.}~\bibnamefont
  {Golestanian}},\ }\href {\doibase 10.1088/1367-2630/ab90d8} {\bibfield
  {journal} {\bibinfo  {journal} {New Journal of Physics}\ }\textbf {\bibinfo
  {volume} {22}},\ \bibinfo {pages} {063045} (\bibinfo {year}
  {2020})}\BibitemShut {NoStop}%
\bibitem [{\citenamefont {{Meng}}\ \emph {et~al.}(2021)\citenamefont {{Meng}},
  \citenamefont {{Matsunaga}}, \citenamefont {{Mahault}},\ and\ \citenamefont
  {{Golestanian}}}]{meng2021magnetic}%
  \BibitemOpen
  \bibfield  {author} {\bibinfo {author} {\bibfnamefont {F.}~\bibnamefont
  {{Meng}}}, \bibinfo {author} {\bibfnamefont {D.}~\bibnamefont {{Matsunaga}}},
  \bibinfo {author} {\bibfnamefont {B.}~\bibnamefont {{Mahault}}}, \ and\
  \bibinfo {author} {\bibfnamefont {R.}~\bibnamefont {{Golestanian}}},\ }\href
  {\doibase 10.1103/PhysRevLett.126.078001} {\bibfield  {journal} {\bibinfo
  {journal} {\prl}\ }\textbf {\bibinfo {volume} {126}},\ \bibinfo {eid}
  {078001} (\bibinfo {year} {2021})}\BibitemShut {NoStop}%
\bibitem [{\citenamefont {Sirota-Madi}\ \emph {et~al.}(2010)\citenamefont
  {Sirota-Madi}, \citenamefont {Olender}, \citenamefont {Helman}, \citenamefont
  {Ingham}, \citenamefont {Brainis}, \citenamefont {Roth}, \citenamefont
  {Hagi}, \citenamefont {Brodsky}, \citenamefont {Leshkowitz}, \citenamefont
  {Galatenko}, \citenamefont {Nikolaev}, \citenamefont {Mugasimangalam},
  \citenamefont {Bransburg-Zabary}, \citenamefont {Gutnick}, \citenamefont
  {Lancet},\ and\ \citenamefont {Ben-Jacob}}]{Sirota-Madi2010genome}%
  \BibitemOpen
  \bibfield  {author} {\bibinfo {author} {\bibfnamefont {A.}~\bibnamefont
  {Sirota-Madi}}, \bibinfo {author} {\bibfnamefont {T.}~\bibnamefont
  {Olender}}, \bibinfo {author} {\bibfnamefont {Y.}~\bibnamefont {Helman}},
  \bibinfo {author} {\bibfnamefont {C.}~\bibnamefont {Ingham}}, \bibinfo
  {author} {\bibfnamefont {I.}~\bibnamefont {Brainis}}, \bibinfo {author}
  {\bibfnamefont {D.}~\bibnamefont {Roth}}, \bibinfo {author} {\bibfnamefont
  {E.}~\bibnamefont {Hagi}}, \bibinfo {author} {\bibfnamefont {L.}~\bibnamefont
  {Brodsky}}, \bibinfo {author} {\bibfnamefont {D.}~\bibnamefont {Leshkowitz}},
  \bibinfo {author} {\bibfnamefont {V.}~\bibnamefont {Galatenko}}, \bibinfo
  {author} {\bibfnamefont {V.}~\bibnamefont {Nikolaev}}, \bibinfo {author}
  {\bibfnamefont {R.~C.}\ \bibnamefont {Mugasimangalam}}, \bibinfo {author}
  {\bibfnamefont {S.}~\bibnamefont {Bransburg-Zabary}}, \bibinfo {author}
  {\bibfnamefont {D.~L.}\ \bibnamefont {Gutnick}}, \bibinfo {author}
  {\bibfnamefont {D.}~\bibnamefont {Lancet}}, \ and\ \bibinfo {author}
  {\bibfnamefont {E.}~\bibnamefont {Ben-Jacob}},\ }\href {\doibase
  10.1186/1471-2164-11-710} {\bibfield  {journal} {\bibinfo  {journal} {BMC
  Genomics}\ }\textbf {\bibinfo {volume} {11}},\ \bibinfo {pages} {710}
  (\bibinfo {year} {2010})}\BibitemShut {NoStop}%
\bibitem [{\citenamefont {Saha}\ \emph {et~al.}(2014)\citenamefont {Saha},
  \citenamefont {Golestanian},\ and\ \citenamefont
  {Ramaswamy}}]{saha2014clusters}%
  \BibitemOpen
  \bibfield  {author} {\bibinfo {author} {\bibfnamefont {S.}~\bibnamefont
  {Saha}}, \bibinfo {author} {\bibfnamefont {R.}~\bibnamefont {Golestanian}}, \
  and\ \bibinfo {author} {\bibfnamefont {S.}~\bibnamefont {Ramaswamy}},\ }\href
  {\doibase 10.1103/PhysRevE.89.062316} {\bibfield  {journal} {\bibinfo
  {journal} {Phys. Rev. E}\ }\textbf {\bibinfo {volume} {89}},\ \bibinfo
  {pages} {062316} (\bibinfo {year} {2014})}\BibitemShut {NoStop}%
\bibitem [{\citenamefont {Pohl}\ and\ \citenamefont
  {Stark}(2014)}]{pohl2014dynamic}%
  \BibitemOpen
  \bibfield  {author} {\bibinfo {author} {\bibfnamefont {O.}~\bibnamefont
  {Pohl}}\ and\ \bibinfo {author} {\bibfnamefont {H.}~\bibnamefont {Stark}},\
  }\href {\doibase 10.1103/PhysRevLett.112.238303} {\bibfield  {journal}
  {\bibinfo  {journal} {Phys. Rev. Lett.}\ }\textbf {\bibinfo {volume} {112}},\
  \bibinfo {pages} {238303} (\bibinfo {year} {2014})}\BibitemShut {NoStop}%
\bibitem [{\citenamefont {Liebchen}\ and\ \citenamefont
  {Levis}(2022)}]{Liebchen_2022}%
  \BibitemOpen
  \bibfield  {author} {\bibinfo {author} {\bibfnamefont {B.}~\bibnamefont
  {Liebchen}}\ and\ \bibinfo {author} {\bibfnamefont {D.}~\bibnamefont
  {Levis}},\ }\href {\doibase 10.1209/0295-5075/ac8f69} {\bibfield  {journal}
  {\bibinfo  {journal} {Europhysics Letters}\ }\textbf {\bibinfo {volume}
  {139}},\ \bibinfo {pages} {67001} (\bibinfo {year} {2022})}\BibitemShut
  {NoStop}%
\bibitem [{\citenamefont {Arumugam}\ and\ \citenamefont
  {Tyagi}(2020)}]{arumugam2020keller}%
  \BibitemOpen
  \bibfield  {author} {\bibinfo {author} {\bibfnamefont {G.}~\bibnamefont
  {Arumugam}}\ and\ \bibinfo {author} {\bibfnamefont {J.}~\bibnamefont
  {Tyagi}},\ }\href {\doibase 10.1007/s10440-020-00374-2} {\bibfield  {journal}
  {\bibinfo  {journal} {Acta Applicandae Mathematicae}\ }\textbf {\bibinfo
  {volume} {171}},\ \bibinfo {pages} {6} (\bibinfo {year} {2020})}\BibitemShut
  {NoStop}%
\end{thebibliography}%

\onecolumngrid

\vspace{12pt}
\noindent\hrulefill \hspace{24pt} {\bf End Matter} \hspace{24pt} \hrulefill
\vspace{12pt}

\twocolumngrid

{\it Appendix A: Boltzmann-Ginzburg-Landau approach.---} 
Here we explicit the truncations of the Boltzmann equation \eqref{eqB} used to obtain the results presented
in Fig.~\ref{fig3}. The self-diffusion integral accounts for the effect of rotational noise on isolated particles:
\begin{equation}
I_{\rm sd}[f] = \lambda \left[-f({\bf r},\theta) + \!\!\int\!\! {\rm d}\theta' f({\bf r},\theta') N_\sigma(\theta-\theta')\right] \,, \nonumber
\end{equation}
where $N_\sigma(\theta)$, in this work, is a zero-mean wrapped Gaussian distribution of variance $\sigma^2$. 
The collision integral reads:
\begin{align} 
I_{\rm co}[f,f] &= -f({\bf r},\theta) \int {\rm d}\theta' K(\theta'-\theta) f({\bf r},\theta') \nonumber \\
+\int\!\! {\rm d}\theta_1& \!\!\int \!\!{\rm d}\theta_2 f({\bf r},\theta_1) K(\theta_2-\theta_1) f({\bf r},\theta_1)
N_\sigma(\theta-\Psi(\theta_1,\theta_2)) \nonumber
\end{align}
where $K(\Delta)=|4r_0v_0\sin(\Delta/2)|$ is the collision kernel between polar particles of radius $r_0$ and speed
$v_0$, and $\Psi(\theta_1,\theta_2)=\tfrac{1}{2}{\rm Arg}[e^{i\theta_1}+e^{i\theta_2}]$ is the ferromagnetic
alignment function.  

The Boltzmann equation \eqref{eqB} expanded over the Fourier angular modes $f_k$ becomes,
using the complex notation  $\triangledown \equiv \partial_x + i\partial_y$ and 
$\triangledown^* \equiv \partial_x - i\partial_y$, the following infinite hierarchy
\begin{align}
\partial_t& f_k +\tfrac{1}{2}v_0(\triangledown^*f_{k+1}+\triangledown f_{k-1}) -ik\omega f_k -D_0\triangledown^* \triangledown f_k =  \nonumber\\
& \lambda (N_k-1)f_k +\sum_{q=-\infty}^{\infty} f_{k-q}f_q [ N_k I_{k,q} - I_{0,q}] \label{hierarchy}
\end{align}
where $N_k=\int_{-\infty}^{\infty}{\rm d}\eta P_\sigma(\eta)\exp(ik\eta)=\exp(-\tfrac{1}{2}\sigma^2k^2)$ 
and 
$$I_{kq}=\tfrac{1}{2\pi} \int_{-\pi}^{\pi}{\rm d}\Delta K(\Delta) e^{-iq\Delta} e^{ik\Delta/2}\;.$$

Applying the scaling ansatz given in the main text, truncating at order $\epsilon^n$, yields $n$ coupled partial
differential equations for the fields $f_0=\rho,f_1,\cdots,f_{n-1}$. Note that the equation for $\rho$ is the classic exact conservation equation, which writes
$\partial_t \rho = - {\rm Re}[ \triangledown^* f_1] +D_0\triangledown\triangledown^*\rho$ 
using our complex notation.\\

{\it Appendix B: Numerical methods.---} 
The microscopic model (\ref{eq:kvm1},\ref{eq:kvm2}) was simulated on square domains of linear size $L$ with periodic boundary conditions, using an explicit Euler-Maruyama scheme with typical timestep $dt=0.1$. 
Total simulation times are of the order of $10^7$-$10^8$ for the largest system sizes.

Whenever the global density or the system size were increased from a configuration with a single condensate (or two in the case of a zero-mean Gaussian distribution of chiralities), particles were added slowly, copying particles taken at random in the densest area, endowing them with a chirality in line with the target distribution $g(\omega)$, at a rate of about 50 new particles per unit time.

Condensates are (slightly) moving, under some residual diffusion. 
Polar packets, moreover, are rotating non-axisymmetric objects.
To perform time-averaging on condensates, we locate their center using a Gaussian smoothing kernel with typical radius larger than their size. This yields smooth density and order fields with a single well-defined peak (in density). 
Once the center located for each instant of interest, time- and azimuthal-averaging 
are easily performed by shifting coordinates to keep a steady center. 
This gives axisymmetric time-averaged fields from which density and order radial profiles are easily extracted.
(For the $p(r)$ order profiles shown in Figs.~\ref{fig2},\ref{fig3} we use the field of the modulus of local polar order.)

In both particle- and continuous-level simulations, $\rho^*$ and $r^*$, for polar packets, are estimated 
by fitting the time-averaged density field
by a Gaussian $\sim\exp(-\tfrac{1}{2} r^2/{r^*}^2)$ over a range $0<r<1.5r^*$. 
For vortices, we fit the region between the gas and the peak density by an exponential $\sim\exp(-r/r_d)$, 
and then calculate $r^*=r_d+r_p$ where $r_p$ is the radius of the maximum density ring, 
which usually is very close to the typical radius of the circle described by a single particle. 

Simulations of the hierarchy \eqref{hierarchy} truncated at $n=5$ presented in Fig.~\ref{fig3} were done with
$\lambda=v_0=2r_0=1$ using a pseudospectral scheme with typical lattice step 1 and time step $2.5\times 10^{-3}$. 
Simulations of the phenomenological equations (\ref{eq:theo1},\ref{eq:theo2}) used a standard finite-difference scheme with lattice step 1 and time step $2\times 10^{-4}$. 
All simulations on square domains of linear size $L$ with periodic boundary conditions, starting from 
an initial Gaussian bump in density.

\begin{figure}[h!]
\includegraphics[width=\columnwidth]{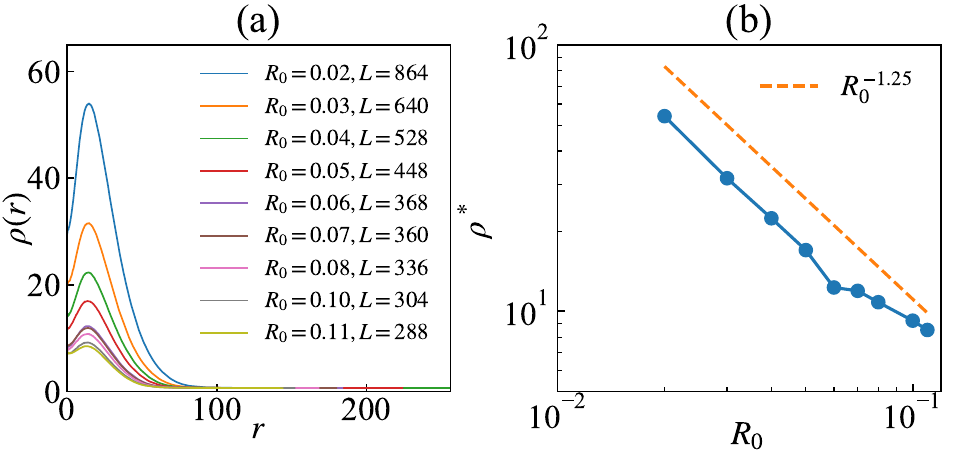}
	\caption{
	Largest size a single vortex condensate can achieve at fixed repulsion strength $R_0$, To explore larger and larger system sizes, we follow the protocol described in Appendix~B, increasing system size by steps $\Delta L=16$, waiting typically $5\times 10^5$ time units to declare the condensate stable. The density profiles shown in (a) are those observed at the last stable size. In (b) we show how the maximal density of these profiles scale with $R_0$. 
	($\kappa=1$, $\omega=1/30$, $D_r=0.085$, Euler timestep $dt=0.01$)
  	}
	\label{figS2}
\end{figure}

{\it Appendix C: Effect of pairwise repulsion.---} 
In Fig.~\ref{figS2}, we show the effect of pairwise, soft, harmonic repulsion on the vortex condensates observed
in the unimodal polar case. Equation~\eqref{eq:kvm1} is now replaced by
\begin{equation} 
\dot{\bf r}_i = {\bf e}(\theta_i) - R_0 \sum_{j\sim i} (d_0-r_{ij}) {\bf e}_{ij}
\end{equation}
where ${\bf e}_{ij}$ is the unit vector parallel to ${\bf r}_j - {\bf r}_i$, $d_0=1$ is chosen for simplicity, and $R_0$ is a measure of the repulsion strength.  
For a given value of the repulsion strength $R_0$, the vortex condensate exists up to 
some size beyond which it typically `explodes' (Fig.~\ref{figS2}(a)). This provides an estimate of the maximal mass and maximal density $\rho^*$ that can be accommodated in a vortex. 
The maximal density scales approximately like $\rho^* \sim R_0^{-1.25}$, and thus diverges as repulsion vanishes, as expected from our conclusion that condensation is boundless in this limit (Fig.~\ref{figS2}(b)).

\end{document}